\documentclass[a4paper,twoside]{article}
\usepackage{amsmath}
\usepackage{amssymb}

\def\Ai{\mathop{\rm Ai}\nolimits}
\def\Bi{\mathop{\rm Bi}\nolimits}
\def\({\left(}
\def\){\right)}
\def\pd#1.#2.#3.{\(\frac{\partial #1}{\partial #2}\)_{#3}}
\def\Or{{\rm O}}
\def\le{\leqslant}

\def\d{{\rm d}}

\begin{document}

\thispagestyle{empty}
\markboth{\hfil The Adiabatic Invariance of the Action Variable in Classical Dynamics\hfil\hfil}{\hfil C.G.~Wells~~and~~S.T.C.~Siklos}
\pagestyle{myheadings}

\centerline{\huge The Adiabatic Invariance of the}
\vskip5pt
\centerline{\huge Action Variable in Classical Dynamics}
\vskip1cm

\centerline{\sc Clive~G.~Wells}
\vskip5pt

\centerline{{\it\small Jesus College, Cambridge CB5 8BL, United Kingdom.}}

\centerline{{\it\small Email address:\/} {\tt cgw11{\rm@}cam.ac.uk}}

\vskip5mm
\centerline{and}
\vskip5mm

\centerline{\sc Stephen~T.~C.~Siklos}
\vskip5pt
\centerline{{\it\small Jesus College, Cambridge CB5 8BL, United Kingdom.}}
\centerline{\it\small and}

\centerline{\it\small Centre for Mathematical Sciences, Wilberforce Road, Cambridge, CB3 0WA, United Kingdom.}

\centerline{{\it\small Email address:\/} {\tt
stcs{\rm@}cam.ac.uk}}

\setcounter{footnote}{0}
\vskip0.3cm

\vskip1cm

\begin{abstract}
We consider one-dimensional classical time-dependent Hamiltonian systems with quasi-periodic
orbits. It is well-known that such systems possess an adiabatic invariant which coincides
with the action variable of the Hamiltonian formalism. We present a new proof of the
adiabatic invariance of this quantity and illustrate our arguments by means of explicit
calculations for the harmonic oscillator.

The new proof makes essential use of the Hamiltonian formalism.
The key step is the introduction of a slowly-varying quantity
closely related to the action variable. This new quantity arises
naturally within the Hamiltonian framework as follows: a canonical
transformation is first performed to convert the system to
action-angle coordinates; then the new quantity is  constructed as
an action integral (effectively a new action variable) using the
new coordinates. The integration required for this construction
provides, in  a natural way, the averaging procedure introduced in
other  proofs, though here it is an average in phase space rather
than over time.

\vskip5mm
\noindent{\sl PACS: 45.05.+x, 45.20.Jj.}
\end{abstract}

\section{Introduction}
\label{sect:intro}

At the Solvay Conference in 1911, Einstein answered a
 question raised by Lorentz with the statement `If the length of a
pendulum
is changed infinitely slowly, its energy remains equal to $h\nu$ if it was
originally $h\nu$'.\footnote{This was apparently first noticed by
Rayleigh in 1902  for the classical pendulum; however, Lorentz's question
related to a pendulum oscillating in a particular {\it quantum\/} state.}
The
expression `adiabatic invariant' came to be used
in this context, and this has stuck over the years,
even though the change of the system has to be slow, rather than
adiabatic in the thermodynamic sense\footnote{
  An adiabatic change in thermodynamics
is one  that happens
with no gain or loss of heat.}
and the quantity in question is not invariant --- and indeed may not
even change  very slowly, as will be discussed further.

Adiabatic invariance is readily understandable at a
superficial level: students  studying classical mechanics
have no difficulty with the idea that, for a simple pendulum,
the quantity $E/\omega$ changes very slowly in some sense when the frequency
$\omega$ is slowly varied. However, the details
are often found more elusive, and with good reason:
the concept is fundamentally
subtle and the proof of invariance is genuinely difficult. The treatment
given in most texts relies on the theory of action-angle variables,
which already involves a significant level of  sophistication;  and
in addition some form of averaging is usually required.
Moreover, many  texts are careless
of details,  omit to state exactly what is being done and do not
reveal the motivation for doing what is done.
There is a danger too, always present when
canonical transformations are used, of losing track of the relation
between the dependent
and
independent variables.

We confine our attention to the case of a one-dimensional system
governed
by a Hamiltonian
\begin{equation}
H(p,q,\lambda(t)),
\label{hamiltonian}
\end{equation}
 where $\lambda$ is slowly
varying in a sense that we will  define precisely in section~\ref{sect:proof}.
The  quantity $I$ defined by
\begin{equation}
I=\frac1{2\pi}\oint p\,\d q
\label{areaintegral}
\end{equation}
is an adiabatic invariant of this system.
The integral is to be  taken round the curve (assumed to be closed) in
phase space given by
$H(p,q,\lambda)=E$ at constant $\lambda$. It represents the area
inside the curve. It is important to realize that this integral is
a purely geometrical construction; it takes no account of the
time evolution of the system.

It is always a good idea to have in mind the example of
the time-varying simple harmonic oscillator with Hamiltonian $H(p,q,
\omega)$
given by
\begin{equation}
H(p,q,\omega(t)) = \frac12(p^2 + \omega^2 (t)q^2).
\label{oscillatorhamiltonian}
\end{equation}
For this system, we may easily calculate the integral in
(\ref{areaintegral}) to obtain
$I= E/\omega$, with the implication that to a good
approximation the energy will
change proportionally to the frequency.

For many simple systems,
it is not even necessary to calculate the integral.
For example, in the case $H=(p^2 +\lambda q^4)/2$,
we have
\begin{equation}
I = \frac 1{2\pi}\oint\sqrt{2E-\lambda  q^4}\, \d q =
\frac1 {\pi}(2E)^{\frac34}\lambda^{-\frac14}
\int_{-1}^1\sqrt{1-x^4}\,\d x
\end{equation}
from which we see that $I$ is a numerical multiple of
$(E^3/\lambda)^{\frac14}$
and that in this case to good approximation the energy varies as $\lambda^{\frac13}$ for slowly varying changes in $\lambda$.

Of course, $I$ is not in general exactly
invariant.\footnote{Interestingly, for the harmonic oscillator, there is an exact
invariant
(called the Ermakov-Lewis invariant --- see \cite{Goodall}
for a useful discussion) which can be written in the form
\[
\tfrac12 \left (
(q/\rho)^2 +(\dot\rho q - \dot q \rho)^2
\right)
\]
where $\rho$ is any solution of $\ddot \rho +\omega^2 \rho =\rho^{-3} $.}
It is necessary therefore to give a precise definition of adiabatic
invariance. Clearly, any quantity such as $I$ (which is essentially
an average over a cycle) will vary slowly, since the system
varies slowly. In order for $I$ to be adiabatic, it must vary more
slowly, in some sense
 than $\lambda$ varies and this means that a proof of
adiabatic invariance must keep careful track of any error terms.

In the next section, we illustrate some of the features of $I$ by
examining a particular adiabatic variation of the harmonic
oscillator. In the third section we discuss strategies for
differentiating $I$ in order to determine its rate of change. In
the fourth section we revisit the harmonic oscillator problem, in
this case we consider a general adiabatic variation and present a
new proof of the adiabatic invariance of $I$. In the fifth
section, we show how this proof can be applied to  more general
systems. Finally in section~6 we summarize our discussion and
present our conclusions.

\section{The time dependence of $I$: harmonic oscillator case}
\label{sect:airy}

In some cases, it is possible to investigate the time
dependence of $I$ by solving Hamilton's equations
and evaluating the area integral (\ref{areaintegral})
explicitly.
It is instructive to look at the simplest case of
the simple harmonic oscillator (\ref{oscillatorhamiltonian}) with
\begin{equation}
\omega^2 = 1+\epsilon t.
\label{specialomega}
\end{equation}
This could be regarded as the
approximation, by Taylor series,
to a more general slowly-varying $\omega(t)$, but for present
purposes that would sacrifice clarity to little purpose: it  would entail
keeping track of an additional set of error terms and
the calculations are messy enough in the special case. For similar reasons we will restrict attention to $\epsilon>0$.
The final expressions would look better
if we had set
$
\omega^2 = \omega^2_0(1+\epsilon t)
$, but the necessary factors  can  easily be
inserted at the end on dimensional considerations.

We will solve Hamilton's equations explicitly for this system in terms
of Airy functions, then calculate the value of $I(t)$ using standard asymptotic
expansions for the Airy functions.

Hamilton's equations are $\dot q=p$, $\dot p=-\omega^2 q$,
leading to
\begin{equation}
\ddot q+(1+\epsilon t)q=0.
\end{equation}
Setting $z= -\epsilon^{-2/3}(1+\epsilon t)$ reduces this
to the Airy equation $q''-zq=0$ with solutions $ \Ai (z)$ and
$\Bi(z)$ (see \cite{Abram} for details). Making use of the fact
that
the Wronskian of $ \Ai (z)$ and
$\Bi(z)$ is $\pi^{-1}$, we can write the solution that satisfies
$q=0$ and $\dot q=1$ at $t=0$ in the form
\begin{equation}
\pi^{-1}
q(t)
=
\Ai\big(-\epsilon^{-\frac23}\omega^2(t)\big)\Bi'\big(-\epsilon^{-\frac23}\big)
-
\Bi\big(-\epsilon^{-\frac23}\omega^2(t)\big)\Ai'\big(-\epsilon^{-\frac23}\big)
.
\end{equation}
For $\epsilon \ll 1$ and $\epsilon t \le 1$, we can apply the
standard asymptotic approximations to obtain
\begin{equation}
q(t) = \omega^{-\frac12} \cos\theta
+ \frac14 \epsilon \omega^{-\frac72}\sin\theta
+\Or(\epsilon^2)\,;
\ \ \ \
p(t) = -\omega ^{\frac12} \sin\theta +  \Or(\epsilon^2),
\end{equation}
where
\begin{equation}
\theta = \frac{2(\omega^3-1)}{3\epsilon}.
\end{equation}
Note that $\theta$, which arises naturally from the asymptotic
expressions, is well behaved in the limit $\epsilon\to0$.
We can now determine the time dependence of $I$:
\begin{equation}
I=H/\omega
=\frac12(p^2/\omega+\omega q^2)=
\frac12+\frac14\epsilon\omega^{-3}\sin\theta\cos\theta
+\Or(\epsilon^2).
\end{equation}
From this result, we can make two important observations:
\begin{itemize}
\item
The time rate of change of $I(t)$  is {\em not\/} smaller than that
of the Hamiltonian; both are first order in the small parameter
$\epsilon$;
\item
$I(t)$ does not grow at a rate proportional to $\epsilon$; rather,
it oscillates and the time-average does not contain a term
proportional to $\epsilon$.
\end{itemize}

\section{The time dependence of $I$: general case}
\label{sect:flaws}

In order to evaluate the integral (\ref{areaintegral}),
we can in principle solve the equation
$H(p,q,t) = E$ at any fixed time $t$, to obtain a solution
 of the form $p=P(E,q, t)$.
Of course, in practice the appropriate value of
$E$ at time $t$  can be obtained explicitly only
by solving Hamilton's equations for the system with appropriate
initial conditions, obtaining a trajectory in phase space
of the from $p=P(t)$, $q=Q(t)$ and setting $E(t) =H(P(t),Q(t),t)$.

At this point, it is instructive to analyse a calculation that is used
 the basis of the proof of adiabatic invariance in a number of
standard
text books.
We start by differentiating under
the integral sign:
\begin{align}
\frac {\d \ }{\d t}\oint P(E,q, t)\,\d q
=&\label{stepone}
\oint \left( \left.\frac{\partial P}{\partial E}\right\vert_{q,t}
\frac{\d E}{\d t}
+
\left.\frac{\partial P}{\partial t}\right\vert_{q,E}
\right)\,\d q\\
=&\label{steptwo}
\oint \left( \left.\frac{\partial P}{\partial E}\right\vert_{q,t}
\left. \frac{\partial H}{\partial t}\right\vert_{p,q}
+
\left.\frac{\partial P}{\partial t}\right\vert_{q,E}
\right)\,\d q\\
=&\label{stepthree}
\oint \left(
\left. \frac{\partial H}{\partial t}\right\vert_{p,q}
\bigg/
\left.\frac{\partial H}{\partial p}\right\vert_{q,t}
+
\left.\frac{\partial P}{\partial t}\right\vert_{q,E}
\right)\,\d q
\\
=&\label{stepfour}
\oint \left(
-\left. \frac{\partial P}{\partial t}\right\vert_{q,E}
+
\left.\frac{\partial P}{\partial t}\right\vert_{q,E}
\right)\,\d q
=0.
\end{align}
This seems to be a good result, but as it stands here
it is wrong. The step that is
incorrect depends on how the integral is to be interpreted and three
different interpretations can be found. It should be said that
the proofs of adiabatic invariance
normally include some sort of averaging which can obfuscate  the deficiencies of the
above argument.

If the integral is interpreted as being  taken
round the  closed curve $H(p,q,t) = E$
(which is how   such integrals will be defined in this paper),
then equation (\ref{steptwo})
is
wrong: the identity $\d E/\d t = \partial H/\partial t$
follows from  Hamilton's equations and Hamilton's equations for the
trajectory of the system whose rate of change of energy is $\d E/\d t$
do not hold on the closed curve. Instead, $\d E/\d t=0$ on this
curve, which leads to a rather unhelpful integral.

If instead the path of integration
is along the trajectory of the system between points at which $q=0$
(say),
then the curve is not closed and attention would have to be paid
to the variation of the endpoints in equation (\ref{stepone});
to put it another way,  the differentiation cannot simply be taken under the
integral sign. As a complicating factor in this case, the integral we are differentiating is not equal to $I$, and extra error terms need to be included to account for this discrepancy.

Finally, if the path of integration is the closed curve $E=H(p,q,t)$
at a fixed time $t$ and then each point on the curve  moves
along the Hamiltonian flow to obtain a closed curve at later times,
then the rate of change of the integral is indeed identically
zero (by Liouville's theorem) but the later curve is no longer
of the form  $E=H(p,q,t)$ (for example,
 it does not remain elliptical in the case of the harmonic oscillator).

The remaining equations are standard results: (\ref{stepthree}) comes
from differentiating $E= H(P(E,q,t),q,t)$ partially with respect to
$E$
and
(\ref{stepfour})
comes from differentiating this same equation partially with respect
to $t$.

Although we do not use the above calculation explicitly in the proof
of adiabatic invariance of $I$, it does provide a useful idea that
underlies
our method of proof. It is relatively easy to calculate $I$ in any
given case, but it is not at all easy to determine its time
dependence. Conversely, there are quantities defined by integrals
closely related to $I$ which are not easy to calculate, but
the time-dependence of which can be determined. The underlying idea
of our proof is to define a quantity $J$ that can be shown to be
close to $I$ and whose time dependence can be determined.

\section{Adiabatic invariance of $I$: the harmonic oscillator case}

The proof of the adiabatic invariance  of $I$ for a general Hamiltonian
presented in the next section is quite difficult to understand at
first because of the number of changes of
variable and the different independent variables used.  This is a
problem common to all such proofs. Instead of starting with the
full proof, it makes sense to see how it works in the more restricted
case of the harmonic oscillator:
\begin{equation}
H(p,q,\tau)=\frac12(p^2+q^2/\tau^2).
\label{oscillatorhamiltonian2}
\end{equation}
We have used $\tau(t)$ as the slowly varying parameter rather than
$\omega$ because it has the dimensions of time.

As mentioned above, the plan is to construct a quantity $J$ which
remains close to $I$ in value as time evolves, but whose time
variation can be determined more easily.

The essence of the proof is to write the system in
action-angle coordinates $(I,\phi)$
with the dynamics governed by the Hamiltonian $K(I,\phi,\tau)$.
Since $K$ is not independent of time we construct an action variable
associated with this new Hamiltonian system.
This new action variable $J$, is an average of $I$ over the
cycle defined by $0\le\phi\le2\pi$.
If the system were independent of time, then $I$ and $J$ would
coincide.
However when $\tau$ is varying they may be shown to differ by an
amount which is $\Or(\dot\tau)$.
The time evolution of $J$ is much slower than that of $I$
and may be determined from the fact that $K$
may be written as a function of $J$ and $\tau$
alone (this follows from the fact that $J$ is the action variable
derived from the Hamiltonian $K$).
In this way we are able to deduce that
$J=\Or(\ddot\tau,\dot\tau^2)$
and the adiabatic invariance of $J$ and hence $I$ may then be established.

We start by defining a new pair of coordinates $(I,\phi)$ by
\begin{equation}
I = \frac12(p^2\tau +q^2/\tau) \equiv H\tau,
\ \ \ \
\phi = \sin^{-1} \frac q {\sqrt{p^2 \tau^2 +q^2}} \equiv\sin^{-1}\frac q
     {\sqrt{2H\tau^2}}.
\label{newcoords1}
\end{equation}
$I$ is the adiabatic invariant defined by the integral
(\ref{areaintegral}).
It may easily be verified that if $p$ and $q$ satisfy Hamilton's
equations
with respect to the Hamiltonian (\ref{oscillatorhamiltonian2}),
then $I$ and $\phi$ satisfy Hamilton's equations with respect to the
Hamiltonian
\begin{equation}
K(I,\phi, \tau) =I (1-\dot\tau \sin\phi\cos\phi)/\tau
.\label{K}
\end{equation}
The coordinates $(I,\phi)$ are therefore canonical\footnote{These coordinates may be
derived
from the type 2 generating function
\[
F(I,q,\tau) =\int_0^q p\,\d q
=\int_0^q\sqrt{2I/\tau-q^2/\tau^2}\,\d q
= I\sin^{-1}(q/\sqrt{2I\tau}) +I(q/\sqrt{2I\tau})
\sqrt{1-q^2/2I\tau}.
\label{F}
\]
}.

Now we define the quantity $J$ by
\begin{equation}
J=\frac1{2\pi}\int_0^{2\pi}I\,\d\phi
\end{equation}
where the path of integration is the closed loop in the $I,\phi$ plane
determined by $K(I,\phi,\tau)$ = constant and $\tau =$ constant.
Note that this is
precisely analogous to the definition (\ref{areaintegral}) of $I$.
Just as $I$ can be written in terms of $p$, $q$ and $\tau$, or
equivalently in terms of $H$ and $\tau$, we can write $J$ in terms
of $I$, $\phi$ and $\tau$, or equivalently in terms of $K$ and $\tau$, using
equation
(\ref{K}):
\begin{equation}
J=\frac{K\tau}{2\pi}\int_0^{2\pi}\frac{\d\phi}{1-\dot\tau\sin\phi\cos\phi}
=K\tau\sqrt{1+\dot\tau^2/4}\;.
\label{K1}
\end{equation}
The value of the above integral can be most conveniently obtained
using an integral round a unit circle in the complex plane.
We can now relate $J$ to $I$ by again using  equation
(\ref{K}):
\begin{equation}
J=I(1-\dot\tau\sin\phi\cos\phi)\sqrt{1+\dot\tau^2/4}.
\label{I}
\end{equation}
Thus
\begin{equation}
|J-I|=\Or(\dot\tau)
\label{IJ}
\end{equation}
which is the key result that ties $I$ to $J$.

Next we need to determine the evolution of $J$.
Differentiating equation (\ref{K1}) gives
\begin{align}
\frac{\d J}{\d t} &= \frac{\d \ }{\d t} \left( K
\tau\sqrt{1+\dot\tau^2/4}\right)
=
 \tau\sqrt{1+\dot\tau^2/4} \frac{\d K}{\d t}
+ K  \frac{\d \ }{\d t}  \left(\tau\sqrt{1+\dot\tau^2/4}\right)\\
&=\tau\sqrt{1+\dot\tau^2/4}
 \left(\frac{\partial K}{\partial t}\right)_{\!\!I,\phi}
+ K  \frac{\d \ }{\d t}  \left(\tau\sqrt{1+\dot\tau^2/4}\right)\\
&=
 \tau\sqrt{1+\dot\tau^2/4}
 \left(\frac{\partial \ }{\partial t}\right)_{\!\!I,\phi}
\left(I (1-\dot\tau \sin\phi\cos\phi)/\tau\right)
+ K  \frac{\d \ }{\d t}  \left(\tau\sqrt{1+\dot\tau^2/4}\right)\\
&= I \left(\frac{\partial \ }{\partial t}\right)_{\!\!I,\phi}
\left( (1-\dot\tau \sin\phi\cos\phi)\sqrt{1+\dot\tau^2/4}\right).
\end{align}
Thus
\begin{equation}
\frac1J \frac{\d J}{\d t} =
\frac{-\ddot\tau \sin\phi\cos\phi}{1-\dot\tau\sin\phi\cos\phi}
+\frac{\dot\tau\ddot\tau}{4+\dot\tau^2}.
\end{equation}

This is the result we want. It shows that, if $\dot\tau $ and
$\ddot\tau$
are small, then $J(t)$ will not vary much from its original value
and nor, in view of (\ref{IJ}), will $I(t)$. The details of this
final step are deferred to the next section, once it has been
determined
exactly what is required from a formal definition
of adiabatic invariance.

\section{Adiabatic invariance of $I$: the general case}
\label{sect:proof}

First we need to adopt a definition of adiabatic invariance.
The most useful one for our purposes is as follows:

Let $T$ be an arbitrary fixed time, for all $\epsilon>0$ consider variations $\tau(t)\equiv\tau(\epsilon,t)$
such that $\dot\tau=\Or(\epsilon)$ and $\ddot\tau=\Or(\epsilon^2)$ (e.g., $\tau(t)=h(\epsilon t)$ for some function $h$). We say that a quantity $I(t)$ is {\em an adiabatic invariant\/} of the dynamical system if for all such variations of $\tau$ we have $|I(t)-I(0)|=\Or(\epsilon)$ for all $0\le t\le T/\epsilon$.\ \footnote{
More formally we mean that $\max\,\{|I(t)-I(0)|:0\le t\le  T/\epsilon\}=\Or(\epsilon)$ for all one parameter adiabatic variations $\tau=\tau(\epsilon,t)$ with $\dot\tau=\Or(\epsilon)$, $\ddot\tau=\Or(\epsilon^2)$ and $I$ evaluated with  $\tau=\tau(\epsilon,t)$.
}

Now we follow the method laid out in the previous section for the
special case of the harmonic oscillator.

From the Hamiltonian $H(p,q,\tau(t))$ we construct new variables $I$
and $\phi$ as follows. First we define $I$:
\begin{equation}
I=\frac1{2\pi}\oint p \,\d q =
\frac{1}{2\pi} \oint P(H, q,\tau) \,\d q
\label{Igeneral}
\end{equation}
where the integral is taken over the curve (assumed to be closed)
$H(p,q,\tau) = $ constant at constant $\tau$.
 As before, $I$ will emerge as a
function of $H$ and $\tau$.
The generating function for the canonical transformation is given by
\begin{equation}
F(I,q,\tau)=\int_0^q p \,\d q',
\end{equation}
where $p$ in the integral is written in terms of $I$, $q'$ (a dummy
variable)
and $\tau$ using (\ref{Igeneral}). Then the angle variable $\phi$ and the new Hamiltonian $K$ are given by
\begin{equation}
\phi=\left(\frac{\partial F}{\partial I}\right)_{q,\tau};
\qquad
K=H  +   \pd F.t.I,q. = H + \dot\tau  \pd F.\tau.I,q.  ,
\label{KH}
\end{equation}
where $H$ is written as a function of $I$ and $\tau$ only.

Now we define $J$ by
\begin{equation}
J(K,\tau)=\frac1{2\pi}\int_0^{2\pi}
I\,\d\phi
\label{Jint}
\end{equation}
where the integral is to be taken round the curve (assumed closed)
in the $(I,\phi)$ plane given by $K= $ constant. If we write $I$ as
a function of $H$ and $\tau$ given by $I(H,\tau)$, we can use
using (\ref{KH}) and a Taylor series expansion to obtain:
\begin{equation}
I(H,\tau) = I\bigg(K-  \dot\tau  \pd F.\tau.{\!\!I},q.,\;\tau\bigg)
= I(K,\tau) -
\dot\tau  \left(\frac{\partial F}{\partial \tau}\right)_{I,q}
\left( \frac{\partial I}{\partial
  H}\right)_{\tau} +\Or(\dot\tau^2)
\end{equation}
where the final partial derivative is to be evaluated at $H=K$.
Substituting into (\ref{Jint}), and remembering that $K$ is constant
in  this integral, gives
\begin{align}
J(K,\tau) &=I(K,\tau)-
\frac{\dot\tau}{2\pi}\left( \frac{\partial I}{\partial H}\right)_{\!\!\tau}
\int_0^{2\pi}
\left(\frac{\partial F}{\partial \tau}\right)_{I,q}\,\d\phi
+\Or(\dot\tau^2)\\
&=
I(H,\tau)+
\dot\tau  \left(\frac{\partial F}{\partial \tau}\right)_{I,q}
\left( \frac{\partial I}{\partial
  H}\right)_{\tau}
-
\frac{\dot\tau}{2\pi}\left( \frac{\partial I}{\partial H}\right)_{\!\!\tau}
\int_0^{2\pi}
\left(\frac{\partial F}{\partial \tau}\right)_{I,q}\,\d\phi
+\Or(\dot\tau^2)\\
&=
I(H,\tau)+
\frac{\dot\tau}{\omega}  \left(\frac{\partial F}{\partial \tau}\right)_{I,q}
-
\frac{\dot\tau}{2\pi\omega}
\int_0^{2\pi}
\left(\frac{\partial F}{\partial \tau}\right)_{I,q}\,\d\phi
+\Or(\dot\tau^2)
\end{align}
where we have written $\omega$ for $\partial H/\partial I$ at fixed $\tau$. Here we assume that $\omega$ is bounded away from zero throughout the motion. This corresponds to the requirement that the system remains in a quasi-periodic state for all times under consideration.
Neither $\omega$ nor
$\left(\frac{\partial F}{\partial \tau}\right)_{I,q}$ depend
explicitly on the small parameter $\dot\tau$, so the first result,
namely at any given time
\begin{equation}
\vert J-I\vert = \Or(\dot\tau)
\label{eq:firstresult}
\end{equation}
is established.

Now we have to investigate the time evolution of $J$.
Using that fact that $J$ depends on $K$ and $\tau$ only,
 we find
\begin{equation}
\frac{\d J}{\d t}
=
\pd J.t.K.+\pd J.K.t.\dot K
=\pd J.t.K.+\pd J.K.t.\pd K.t.I,\phi.=\pd J.t.I,\phi.
\end{equation}
so
\begin{equation}
\frac{\d J}{\d t}=
\pd.t.\phi,I.
\(I+\frac{\dot\tau}{\omega}
\pd F.\tau.I,q.-\frac{\dot\tau}
{2\pi\omega}\int_0^{2\pi}\pd F.\tau.I,q.\,\d\phi
+\Or(\dot\tau^2)\)
=\Or(\ddot\tau,\dot\tau^2)
\end{equation}
which establishes the second result.

If the variation in $\tau$ is adiabatically slow, so that $\Or(\ddot\tau,\dot\tau^2)=\Or(\epsilon^2)$, then
by integrating $\d J/\d t$ from $0$ to $t\in[0,T/\epsilon]$ we deduce that $|J(t)-J(0)|=\Or(\epsilon)$. Our first result, equation (\ref{eq:firstresult}), then implies that $|I(t)-I(0)|=\Or(\epsilon)$, i.e., $I$ is an adiabatic invariant.

\section{Conclusion}

The theory of adiabatic invariants is one of the more confusing
aspects of Hamiltonian mechanics in undergraduate courses on
classical dynamics. An intuitive understanding of the concept is
often all that is given, and the subtleties ignored. For instance,
as we point out, it is {\em not\/} true to say that an adiabatic
invariant varies much more slowly than the slowly-varying
parameter. As we discussed in section~\ref{sect:airy}, the action
variable typically has a rate of change comparable with that of
the slowly-varying parameter. However, over suitably defined
extended periods of time, the change in the adiabatic quantity is
much less than the typical change in the quantity being varied. It
is in this rather precise sense that the quantity is `invariant'.

The subtlety in formulating an adequate definition of adiabatic
invariance means that a proof of the adiabatic invariance of the action variable,
for instance, will of necessity require a certain level of sophistication.
In section~\ref{sect:flaws} we pointed out some of the problems
encountered in attempts to prove the result directly, i.e.,
without recourse to the more advanced theory of canonical
transformations and action-angle variables. Frequently attempts to prove the
adiabatic invariance of the action variable introduce a time averaging procedure;
the average being taken over a time scale over which the system is
approximately periodic. One may then be tempted to replace the quantity $I$,
which is a geometrical construction given by an integral over a region
in  phase space bounded by a certain constant energy contour,
with an approximation to $I$ based on the trajectory of the particle in question,
a dynamical construction requiring the use of Hamilton's equations.
This would make no difference  if the  system were independent of time, but in the
time dependent case,
one is then left with the problem of keeping track of the
all the errors made in using these approximations.

In contrast, our new proof constructs $J$, a second action
variable which has the natural interpretation as the phase space
average of $I$ over all points on the constant energy contour, or
more precisely, over the angle variable. This is the analogue of
the time-averaging procedure normally used. Perhaps naturally, in
light of the simplicity with which this fits into the formalism of
Hamiltonian mechanics, this leads to a far clearer proof of the
adiabatic invariance of action variable where error terms can be
easily tracked at every step of the proof.

We have used the harmonic oscillator to illustrate our proof and
to emphasize the subtleties that arise in the theory. For the
simplest of variations Hamilton's equations may be solved exactly
in terms of Airy functions. The asymptotic properties of these
functions are well-known and lead to a concrete example of the
issues we have discussed in this paper. The harmonic oscillator
with arbitrary adiabatic variation is also simple enough to allow
for all of the functions introduced in our general proof to be
written down in closed form. This provides a useful touchstone for
readers following the general proof.

\end{document}